\documentclass[aps,prl,reprint]{revtex4-2} 
\usepackage{graphicx}
\usepackage{dcolumn}
\usepackage{bm}
\usepackage{latexsym}
\usepackage{amssymb}
\usepackage{amsmath}
\usepackage{times}
\usepackage{caption}
\usepackage{subcaption}
\usepackage{lipsum}
\usepackage{listings}
\usepackage{adjustbox}
\usepackage{float}
\usepackage{xcolor}
\usepackage{enumitem}
\usepackage[normalem]{ulem} 
\usepackage{etoolbox}

\begin{document}

    \title{Extraction of Moment Closures for Strongly Non-Equilibrium Flows via Machine Learning} 
    \author{Hang Song}
    \email{song@acom.rwth-aachen.de}
    \affiliation{Applied and Computational Mathematics, RWTH Aachen University, 52062 Aachen, Germany}

    \author{Satyvir Singh}
    \email{singh@acom.rwth-aachen.de}
    \affiliation{Applied and Computational Mathematics, RWTH Aachen University, 52062 Aachen, Germany}
 
    \author{Semih Cayci}
    \email{cayci@mathc.rwth-aachen.de}
    \affiliation{Mathematics of Machine Learning, RWTH Aachen University, 52062 Aachen, Germany}
 
    \author{Manuel Torrilhon}
    \email{mt@acom.rwth-aachen.de}
    \affiliation{Applied and Computational Mathematics, RWTH Aachen University, 52062 Aachen, Germany}

\begin{abstract}
We introduce a machine learning framework for moment-equation modeling of rarefied gas flows, addressing strongly non-equilibrium conditions inaccessible to conventional computational fluid dynamics. Our approach utilizes high-order moments and collision integrals—highly sensitive to non-equilibrium effects—as key predictive variables. Training datasets are created from one-dimensional steady shock simulations, and a methodology of computing collision integrals is developed. By learning thermodynamically consistent closures directly from DSMC data, our R13-ML model, combined with a discontinuous Galerkin solver for the transfer equations of moments, preserves physical structure and accurately predicts normal shock structures and generalizes to hypersonic and some unsteady, one-dimensional wave scenarios. This work bridges machine learning with continuum mechanics, offering a road map for high-fidelity aerothermal predictions in next-generation supersonic vehicles. 
\end{abstract}
\maketitle
\noindent\textbf{Introduction} Non-equilibrium gas dynamics plays a fundamental role in hypersonic aerothermodynamics, micro/nanoscale flows, and vacuum systems, where the breakdown of continuum assumptions renders conventional Navier-Stokes equations inadequate \cite{anderson2006hypersonic,Park1990Nonequilibrium, Pirozzoli2011Numerical}. The degree of rarefaction, characterized by the Knudsen number $\text{Kn}=\lambda/L$, leads to significant deviations from equilibrium at large $\text{Kn}$ due to reduced collisionality, whenever the mean free path $\lambda$ reaches similar magnitudes as the macroscopic length scale $L$.

Current modeling approaches face fundamental challenges. While the direct simulation Monte Carlo (DSMC) method provides high-fidelity resolution of multiscale physics, its computational cost becomes prohibitive in slip and transition regimes \cite{bird1994molecular}. Moment methods offer a promising bridge between kinetic and continuum descriptions \cite{grad1949kinetic}, but classical closures such as Grad 13-moment equations (G13) or regularized 13 moment equations (R13) fail under strong non-equilibrium conditions, primarily due to their inability to accurately represent high-order moments and capture complex collision dynamics especially in the nonlinear regime at high Mach numbers \cite{Torrilhon2016Modeling, Torrilhon2004Regularized}.

Recent advances in machine learning (ML) have opened new avenues for fluid dynamics research, with applications spanning turbulence modeling, rarefied gas transport, and molecular-scale simulations \cite{Brunton2020, Lei2023Machine, Zhao2020RANS, Nejad2021Modeling, Shan2023Turbulence, Kou2022Koopman, Valentini2020NN, Huang2023DMC, Ayyaswamy2012DSMC, li2023learning, mouhot2006fast}. Although neural network architectures have been proposed to approximate the Boltzmann collision integral and extend moment closures \cite{han2019uniformly,xiao2023relaxnet, garg2024neural}, accurate and robust modeling of strongly non-equilibrium regimes remains open.

In this work, we present a machine learning-enhanced moment framework that closely combines moment transfer equations from first principles with data-driven reconstructions of high-order constitutive relations. First, we construct a comprehensive DSMC-generated database encompassing equilibrium to strongly non-equilibrium states through careful normalization and data augmentation techniques. Fully connected neural networks (FCNNs) \cite{nair2023deep} are then embedded within the discontinuous Galerkin spectral element method (DGSEM) solver {\tt Trixi.jl} \cite{Schlottke2021purely} for the moment transfer equations to provide dynamic closure updates for high-order moments and collision integrals. The resulting R13-ML model demonstrates robust predictive capabilities for hypersonic and strongly non-equilibrium flows which generalize from steady into unsteady processes, establishing a new paradigm for ML-enhanced kinetic-fluid modeling.

\noindent\textbf{Physical Model} The theoretical foundation of our approach rests on the Boltzmann equation ${\partial f}/{\partial t} + \bm{c} \cdot \nabla_{\mathbf{x}} f = S[f]$, which provides a complete microscopic description of dilute gas dynamics through a phase space distribution function  \( f(\mathbf{x}, \bm{c}, t) \) \cite{boltzmann1872kinetic}. \( S[f] \) is the collision operator. Macroscopic variables for a monatomic gas with particle mass $m$, density ($\rho= m\int f dc$), momentum ($\rho \mathbf{v}=m\int \bm{c} \, f dc$), internal energy ($\rho e=m\int \frac{1}{2}|\bm{C}|^2 \, f dc$), with thermal velocity $\bm{C}=\bm{c}-\mathbf{v}$, are defined by velocity moments. The transfer equations of these moments follow from the integration of the Boltzmann equation and reveal a closure problem for the shear stress ($\sigma_{ij}= m\int C_{\langle i} C_{j \rangle} \,f$) and the heat flux ($q_i= m\int \frac{1}{2}C^2 C_i \, f$). The classical closure of the Navier-Stokes-Fourier (NSF) theory puts them proportional to the gradient of velocity and the gradient of temperature, respectively.

However, NSF becomes invalid due to the breakdown of the constitutive relations in rarefied flow ($\text{Kn} \gtrsim 0.01 \textnormal{-} 0.1$). The moment approximation theory \cite{Struchtrup2007} adds stress tensor and heat flux as independent variables to be solved as part of the fluid-dynamic system. The additional transfer equations again reveal a closure problem for higher order moments which are written as $m_{ijk}=m\int C_{\langle i} C_{j} C_{k\rangle} \,(f-f_\text{G13}) dc$ and $R_{ij}=m\int C^2 C_i C_j \, (f-f_\text{G13}) dc$ as deviations from the Grad-13-moment theory \cite{Struchtrup2003Regularization}. The so-called R13-closure gives explicit expressions for these moments
\begin{align}
 m_{ijk}^{\text{R13}}\!=&\!-2\mu\Bigl[
 \tfrac{1}{\rho}\tfrac{\partial \sigma_{\langle ij}}{\partial x_{k \rangle}}
 \!-\!\tfrac{\sigma_{\langle ij}}{\rho}\tfrac{\partial \text{ln}\rho}{ \partial x_{k\rangle}}
 \!+\!\tfrac{4}{5}\tfrac{q_{\langle i}}{p}\tfrac{\partial v_j}{\partial x_{k \rangle}}
 \Bigl]
 \label{m_closure}
 \\
 R_{ij}^{\text{R13}}\!=&
 -\tfrac{24}{5}\mu \Bigl[ \tfrac{1}{\rho}\tfrac{\partial q_{\langle i}}{\partial x_{j\rangle}} 
 \!+\! \tfrac{Rq_{\langle i}}{p} \tfrac{\partial T}{\partial x_{j\rangle}} 
 \!-\! \tfrac{q_{\langle i}}{\rho} \tfrac{\partial \ln \rho}{\partial x_{j\rangle}} 
  \label{R_closure} \\
  &+ \tfrac{5}{7} \Bigl( \tfrac{\sigma_{k\langle i}}{\rho} \tfrac{\partial v_{j\rangle}}{\partial x_k} \!+\! \tfrac{\sigma_{k\langle i}}{\rho} \tfrac{\partial v_{k}}{\partial x_{j\rangle}} \!-\! \tfrac{2}{3} \tfrac{\sigma_{ij}}{\rho} \tfrac{\partial v_{k}}{\partial x_k} \Bigl)\Bigl]\notag
 \\
 &-4 \mu\Bigl[ \tfrac{1}{\rho}\tfrac{\partial q_k}{\partial x_k}
\!+\! \tfrac{5}{2}\tfrac{R q_k}{p} \tfrac{\partial T}{\partial x_k}
\!-\! \tfrac{q_k}{\rho}\tfrac{\partial \ln \rho}{\partial x_k}
\!+\! \tfrac{\sigma_{kl}}{\rho}\tfrac{\partial v_k}{\partial x_l}\Bigl]\notag
 \end{align}
 based on gradients of stress and heat flux but also on equilibrium moments in a nonlinear way. The final R13 system \cite{Struchtrup2007,Torrilhon2016Modeling} of evolution equations can be written 1D as ${\partial \mathbf{U}}/{\partial t}+ {\partial \mathbf{F}[\mathbf{U}]}/{\partial x}= \mathbf{Q}(\mathbf{U})$, where $\mathbf{U}$ represent the moment vector, $F[\mathbf{U}]$ the fluxes including gradient terms and $\mathbf{Q}(\mathbf{U})$ the moments of the collision operator \( S[f] \), which includes the production of stress $Q_{ij}=m\int C_i C_j \,S[f] dc$ and heat flux $Q_i=m\int C^2 C_i \,S[f] dc$ based on the R13 theory.

The connection between microscopic kinetics and macroscopic continuum mechanics is established through rigorous moment-integration of the Boltzmann equation. Within this R13 framework, the accuracy of rarefied flow simulations fundamentally depends on three cornerstones: (i) inclusion of the conservation laws and cross coupling of intermediate moments, (ii) physically consistent closures for high-order moments, and (iii) mathematically precise formulations of collision integrals. These  requirements collectively determine the solution fidelity when using moment equations in the simulation of rarefied flow situations and will be retained in our machine-learning model.

In weakly nonequilibrium regimes, closures based on linear theory like R13 \eqref{m_closure}/\eqref{R_closure} and simplified collision models, e.g., BGK \cite{BGK1954Model},  remain valid. However, under strong nonequilibrium conditions \eqref{m_closure}/\eqref{R_closure} do not hold anymore, but nonlinear moment relations and collision integrals become analytically intractable, necessitating expensive particle methods, like DSMC. DSMC naturally resolves molecular collisions and rarefaction effects, providing reliable high-order moments and collision terms, though at high computational cost. 

In this paper we will replace the R13 closure for both the higher order moments and the collision integrals by a carefully trained machine-learning model $m_{ijk}^{\text{ML}}$, $R_{ij}^{\text{ML}}$, as well as $Q_{ij}^{\text{ML}}$ and $Q_i^{\text{ML}}$.

\noindent\textbf{Data Sets and Scaling }A shock structure manifests as a narrow stationary transitional zone between a supersonic upstream and subsonic downstream flow. We construct training datasets from one-dimensional monatomic argon shock waves, spanning Mach numbers $\text{Ma}=1.2\textnormal{-}8.0$ in $0.4$ increments. Each of the 18 steady cases is discretized into 800 grid points with complete flow variables and high-order moments sampled through the molecular thermal velocity. While moments are easily accessible in DSMC determining the collision integral \(\int  (\cdot)S[f]dc \) presents a formidable challenge. Leveraging the steady-flow condition \( \partial \mathbf{U} / \partial t = 0 \), this study determines the source terms of moment equations \(\mathbf{Q}(\mathbf{U}) \) via the steady-state relation ${\partial \mathbf{F}(\mathbf{U})}/{\partial x} = \mathbf{Q}(\mathbf{U})$. Since the fluxes are essentially given by moments, the collison integral is extracted from training dataset 
\begin{equation}
\mathbf{Q}(\mathbf{U})\big|_{x_i}=\tfrac{\partial \mathbf{F}}{\partial x} \big|_{x_i} \approx \tfrac{\mathbf{F}_{i+1} - \mathbf{F}_{i-1}}{2 \Delta x}
\label{collision integral}
\end{equation}
by discretizing the 1D spatial derivative of the flux. To enhance data density while mitigating DSMC noise, we implement polynomial interpolation between adjacent points and synthetic augmentation via flow reversal. This strategy effectively doubles usable samples and ensures robust datasets for machine learning closure extraction.

To ensure consistent scaling across equilibrium and nonequilibrium regimes, all variables are nondimensionalized using density $\rho$, reference temperature $T_r$ and temperature-dependent Newtonian speed of sound $c_{s}(T)=\sqrt{R T}$ as well as mean free path $\lambda(\rho,T)={m}/({\sqrt{2}\,\rho \pi d^{2}({T_{r}}/ {T})^{\omega-0.5}})$,
where $d$ is the molecular diameter, and $\omega$ the viscosity index of the gas. The reference speed of sound is denoted by $a=c_s(T_r)$.

The normalized input variables are taken from \eqref{m_closure} and \eqref{R_closure}, including local moments and their derivatives
\begin{align}
&\begin{array}{l}\displaystyle
\frac{T}{T_{r}},\;\frac{p}{\rho a^{2}},\;\frac{\partial_{x}\rho}{\rho/\lambda},\;\frac{\partial_{x}v}{a/\lambda},\;\frac{\partial_{x}T}{T_{r}/\lambda},\;\frac{\partial_{x}p}{\rho a^{2}/\lambda},\\ \displaystyle
\frac{\sigma_{xx}}{\rho a^{2}},\;\frac{q_{x}}{\rho a^{3}},\;\frac{\partial_{x}\sigma_{xx}}{\rho a^{2}/\lambda},\;\frac{\partial_{x}q_{x}}{\rho a^{3}/\lambda},
\end{array}\label{input}
\end{align}
while Galilean invariance is enforced by excluding macroscopic velocities. 
The normalized 1D outputs are
\begin{equation}
\frac{m_{xxx}^{\text{ML}}}{\kappa\rho a^{2}c_{s}},\quad
\frac{R_{xx}^{\text{ML}}}{\kappa\rho a^{3}c_{s}},\quad
\frac{Q_{xx}^{\text{ML}}}{\rho a^{2}c_{s}/(\kappa\lambda)},\quad
\frac{Q_{x}^{\text{ML}}}{\rho a^{3}c_{s}/(\kappa\lambda)}.\label{output}
\end{equation}
in which $\kappa=5\sqrt{2\pi}(\alpha+1)(\alpha+2)/(4\alpha (5-2\omega)(7-2\omega))$ with $\alpha$ the VSS model parameter controlling the collision deflection angle \cite{bird1994molecular}. With this data set we only train the dependency of output \eqref{output} on input \eqref{input}, not for the spatial shape of shock wave profiles. Note that besides reference values, density- and temperature-dependent sound speed and mean free paths are used in the scaling. This preserves rarefaction effects and ensures stable machine-learning training.

\begin{figure*}[t]
    \centering
    \includegraphics[width=1.0\textwidth]{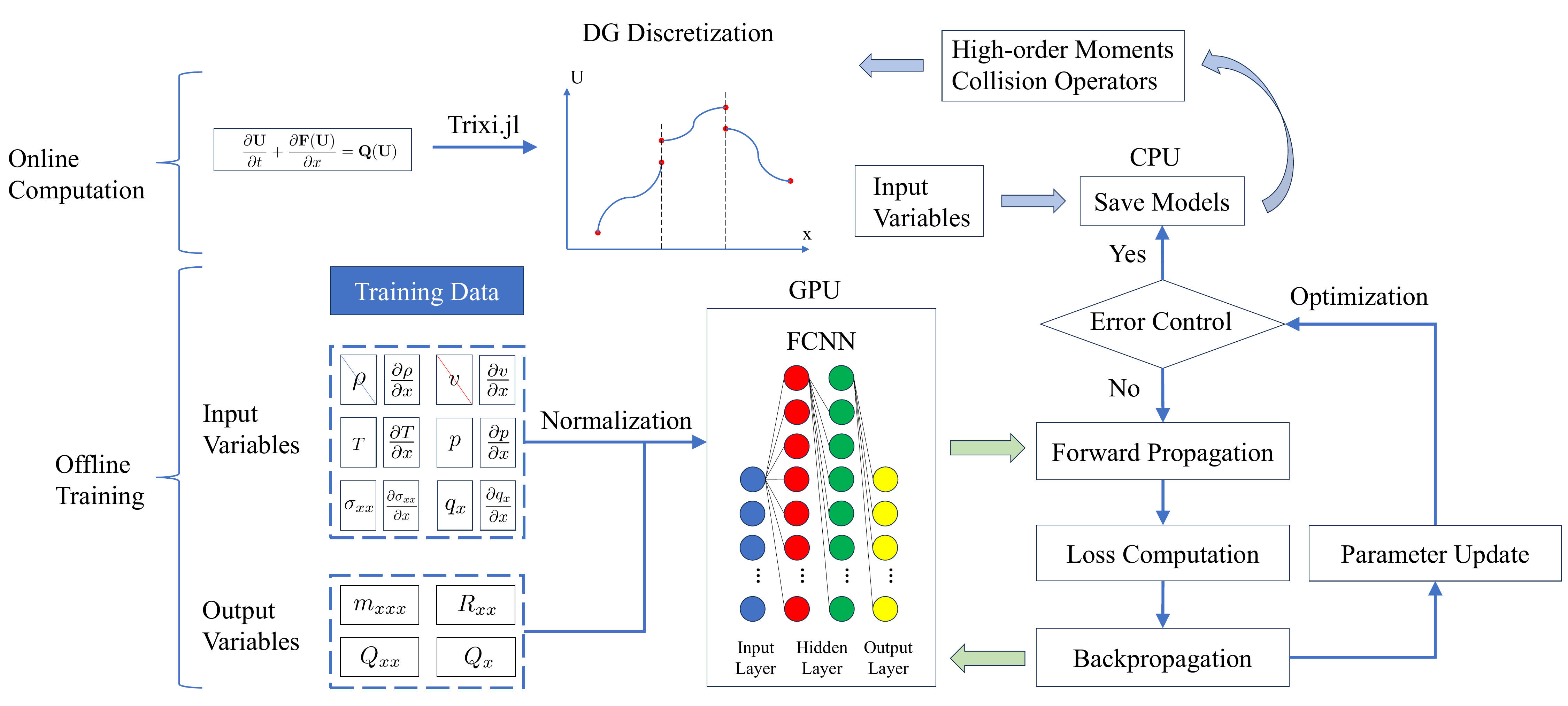}
    \caption{Overview of the R13-ML-model based on an FCNN with offline training and physics-based online computation \cite{github_R13ML}}
    \label{fig:1}
\end{figure*}

\noindent\textbf{ML Model }As an offline preparation the normalized training data derived from DSMC simulations is fed into a fully connected neural network (FCNN) to establish nonlinear mappings between input variables and corresponding output responses, as schematically illustrated in Fig. \ref{fig:1}. To enhance efficiency, the network training is accelerated through GPU-based parallel processing. In this research, each output variable is modeled by an independently trained network to prevent interference. The FCNN architecture consists of an input layer, six hidden layers ($128 \to 64 \to 64 \to 64 \to 64 \to 64$) with softplus activation functions, and an output layer. The trained neural network is subsequently deployed on CPU architectures for seamless integration with the {\tt Trixi.jl} computational framework. This online computation is using a physics-based discretization of the partial differential equations given by the moment approximations of Boltzmann equation \cite{github_R13ML}, hence ensuring preservation of physical properties like conservation laws, etc. 

\begin{figure}[b]
    \centering
    \includegraphics[width=0.5\textwidth]{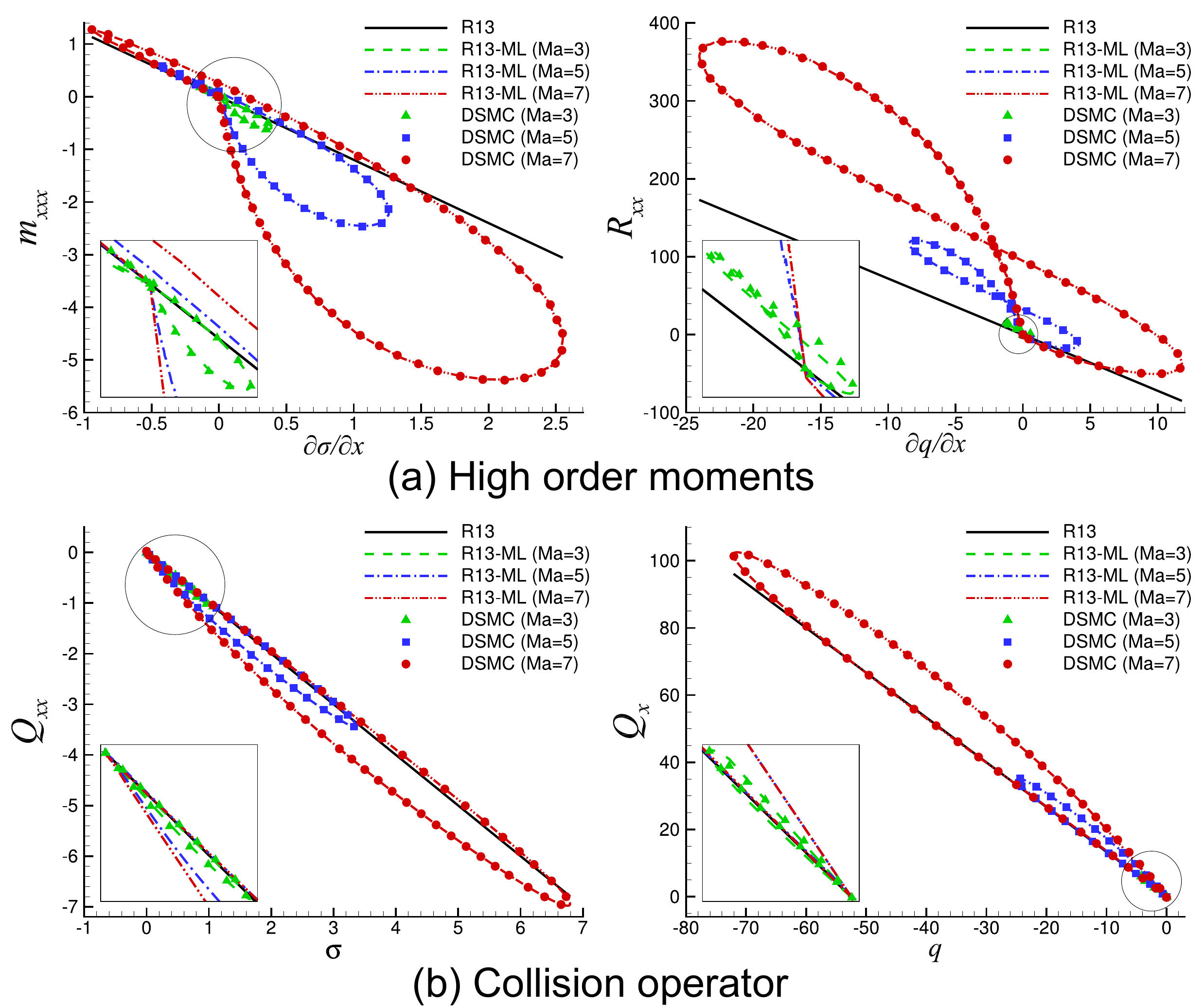}
    \caption{Performance of the R13-ML model on test data: Moment dependencies in 1D shocks at $\text{Ma}=3,5,7$}
    \label{fig:2}
\end{figure}

\noindent\textbf{Results } In agreement with the training data this computational study employs molecular argon as the working medium for simulating non-equilibrium flow phenomena. Initial conditions specify a pre-shock mean free path of 1 mm at 273.15 K, with reference parameters derived from \cite{bird1994molecular}. Fig. \ref{fig:2} shows excellent agreement for the behavior of high-order moments and collision integrals between R13-ML predictions and DSMC references for test dataset. Closer looks reveals the deviation trend: as the Mach number increases, the discrepancies in high-order moments between the DSMC and the linear-theory-based R13 baseline model progressively intensify. However, the R13-ML's accuracy in capturing complex closure behaviour within the training regime is clearly demonstrated. 

\begin{figure*}[t]
    \centering
    \includegraphics[width=1.0\textwidth]{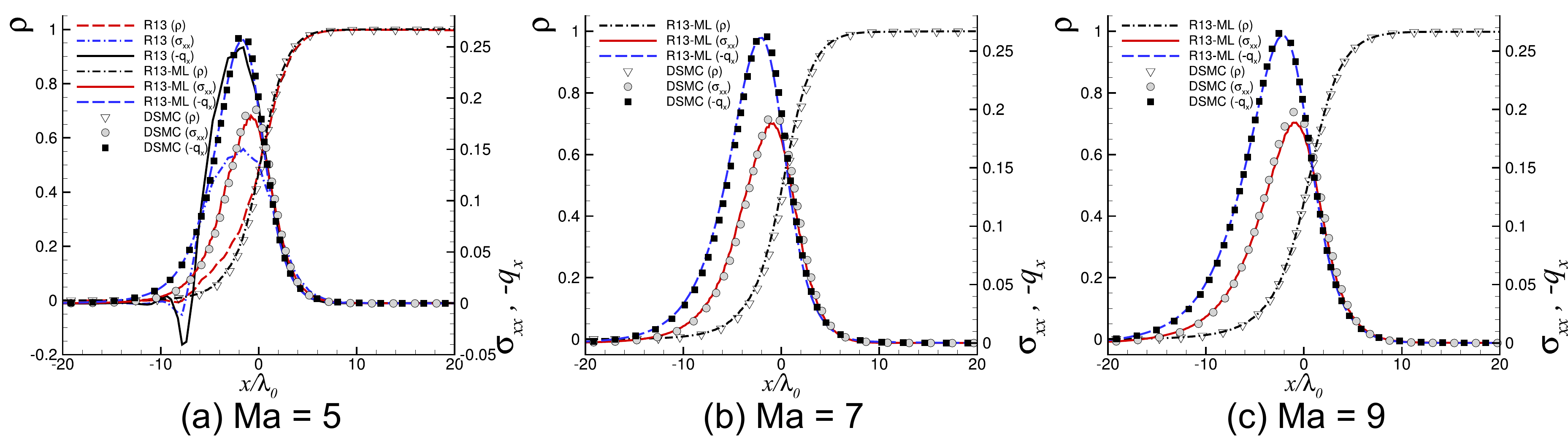}
    \caption{One-dimensional shock structure problem: solution comparison for normalized stress ($\sigma$) and normalized heat flux ($q$) at $\text{Ma}=5, 7, 9$. Excellent agreement also for $\text{Ma}=9$, which is outside the trainings set.}
    \label{fig:3}
\end{figure*}

The actual predictive capability of the FCNN model becomes apparent after implementing the R13-ML model for the closure variables \eqref{output} into the moment equations discretization based on the DGSEM algorithm provided in {\tt Trixi.jl}. The R13-ML model demonstrates exceptional performance in simulating one-dimensional shock waves at $\text{Ma}=5,7$, shown in Fig. \ref{fig:3}. Its extrapolation capacity is demonstrated by good agreement for higher Mach numbers which were not included in the training data. The figure shows $\text{Ma}=9$. 

\noindent\textbf{Transient Generalization }Critically, the model also demonstrates robust generalization to strongly non-equilibrium, unsteady flow fields. 
\begin{figure*}[t]
    \centering
    \includegraphics[width=1.0\textwidth]{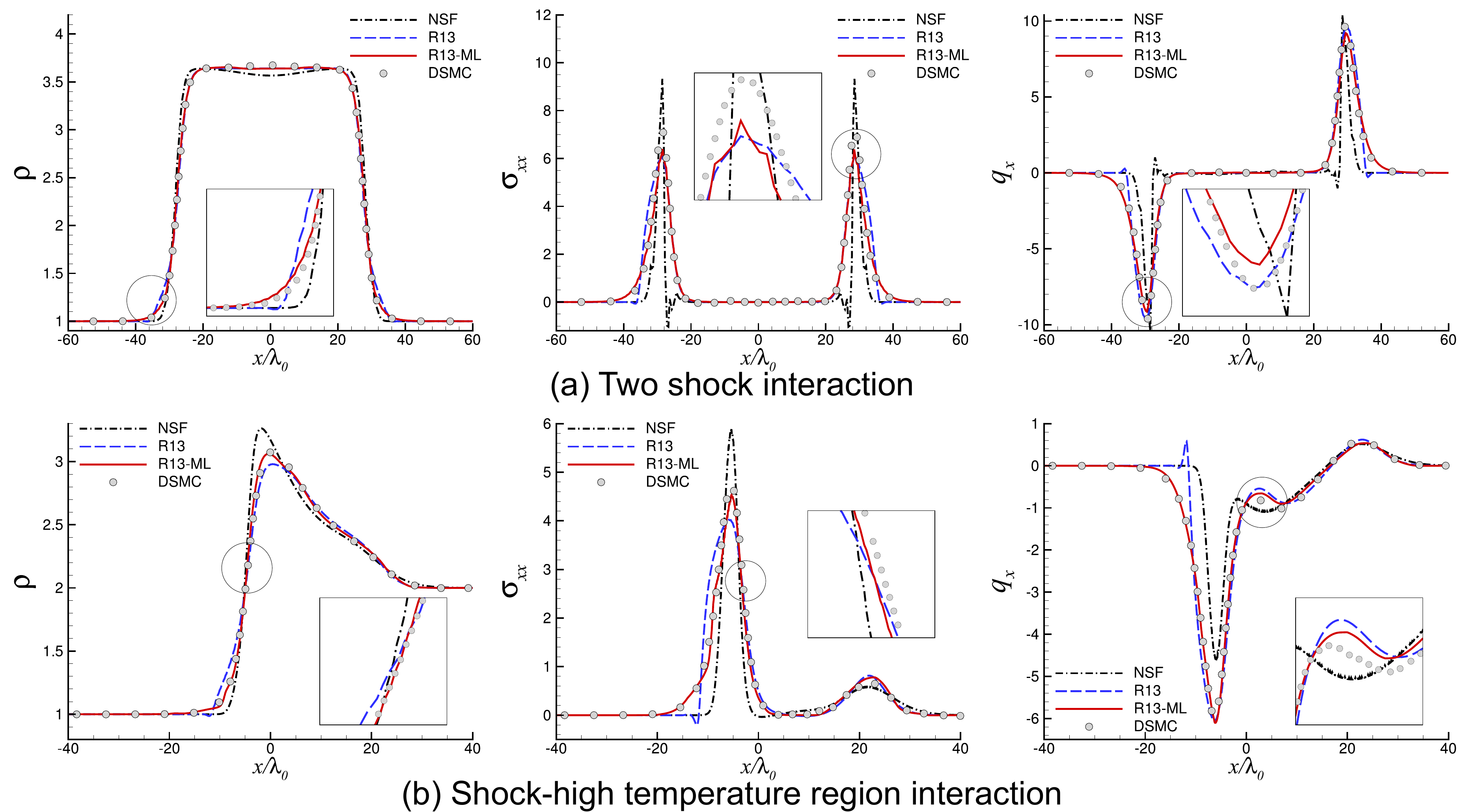}
    \caption{Simulation of unsteady flow: (a) two shock interaction at $t=22.830\tau_0$ after the collision of two shock waves. (b) shock-high temperature region interaction at $t=7.610\tau_0$ after impingement of a stream on a high temperature region.}
    \label{fig:4}
\end{figure*}
Fig. \ref{fig:4}(a) displays the transient flow field resulting from the interaction of two approximately Mach 4 argon shocks at $t = 22.830 \tau_0$ ($T = 273$ K, mean free path $\lambda_0= 1$ mm, mean collision time $\tau_0=2.628\mu\text{s}$). The initial condition is given by
\begin{equation*}
	\begin{aligned}
		(\rho, v, p, \sigma_{xx}, q_x) &=
		\begin{cases}
			(1, 1, 1, 0, 0), & \text{if } x \leq 0, \\
			(1, -1, 1, 0, 0), & \text{if } x > 0.
		\end{cases} \\
	\end{aligned}
	\label{two shock interaction}
\end{equation*}
The R13-ML model exhibits significantly higher accuracy than both the NSF and standard R13 equations. It shows excellent agreement with DSMC results in terms of density, while only minor deviations are observed in the peaks of stress and heat flux, with relative errors below 5\%. In comparison, NSF and R13 yield substantial errors across all flow variables within the bilateral wavefront regions.

Fig. \ref{fig:4}(b) depicts another transient flow field at $t=7.610\tau_0$, after impingement of a argon stream (initial temperature: 273.15 K, mean free path $\lambda_0=1$ mm, mean free flight time $\tau_0=2.628\mu\text{s}$, $\text{Ma}\approx4$) upon a predefined high-temperature region with doubled ambient density. The initial interface separating the domain is given by
\begin{equation*}
	\begin{aligned}
		(\rho, v, p, \sigma_{xx}, q_x) &=
		\begin{cases}
			(1, 1, 1, 0, 0), & \text{if } x \leq 0, \\
			(2, 0, 20, 0, 0), & \text{if } x > 0.
		\end{cases} \\
	\end{aligned}
	\label{shock-high temperature region interaction}
\end{equation*}
The R13-ML model again demonstrates markedly superior accuracy compared to both NSF and standard R13 equations. It achieves near-perfect agreement with DSMC results in density and stress, while heat flux exhibits a slight deviation in the region after the expansion wave ($3 < x/\lambda_0 < 10$). In contrast, NSF severely underestimates the wave thickness, while R13 fails to reproduce the wavefront region.

\noindent\textbf{Conclusion }This work addresses the challenge of simulating hypersonic rarefied flows, where strong non-equilibrium effects invalidate classical constitutive relations and moment closures, rendering NSF and moment closures based on linear theory ineffective. Using DSMC data across Mach numbers $1.2$–$8.0$, we constructed a comprehensive dataset of dependencies of high-order moments and collision integrals on local conditions. From this, we developed the R13-ML model via physics-informed normalization and scaling, and neural-network-based extraction of closure relations. Coupled with a numerical method to solve the underlying physical moment equations, the model achieves accurate and efficient simulations of non-equilibrium shocks.

The R13-ML closure not only reproduces training cases but also extrapolates robustly to hypersonic shock wave, and even unsteady, transient flows. This study demonstrates that -- when used in the right place of physical descriptions -- machine learning can enable moment equations effectively for high-fidelity rarefied hypersonic simulations. In the future R13-ML will be extended to multiple space dimensions using appropriate data sets.

\noindent\textbf{Acknowledgments }H.S. would like to acknowledge funding through the German National Network for High-Performance Computing (NHR). S.S. and M.T. acknowledge the support provided by the German Research Foundation within the research unit DFG–FOR5409.

\noindent\textbf{Data and code availability }The trainings data, pre-trained closure models and the numerical solver are publically available on GitHub \cite{github_R13ML}.

\end{document}